\documentclass[twocolumn,showpacs,preprintnumbers,amsmath,amssymb]{revtex4}
\usepackage{epsfig}
\usepackage{amsmath}

\begin{document}

\title{Synchronization of Non-Linear Random Walk Dynamics 
in Complex Networks}

\author{Luciano da Fontoura Costa}
\affiliation{Institute of Physics at S\~ao Carlos, University of
S\~ao Paulo, PO Box 369, S\~ao Carlos, S\~ao Paulo, 13560-970 Brazil}

\date{9th Jan 2008}

\begin{abstract}
This work addresses synchronization in transient, non-linear
stochastic dynamics corresponding to accesses performed by
self-avoiding walks originating at each node of a complex network.
More specifically, the synchronizability of accesses incoming from
other nodes has been considered and quantified in terms of the entropy
of the mean periods of access, being closely associated to the
efficiency of access delivery to each node.  The concept of
synchronous support of a node $i$ has also been suggested as
corresponding to the nodes which contribute the most for the
synchronization of the accesses to $i$.  These concepts have been
applied to the analysis of 6 networks of different types, leading to
markedly smaller synchronizability being obtained for the
Watts-Strogatz and a geographical models.  The more uniform
synchronizabilities were identified for the Watts-Strogatz and
path-regular structures. Varying degrees of correlations were found
between the synchronizability and the degree or outward accessibility
of the nodes.  The synchronous support of a node $i$ has been found to
present diverse structure, including nodes which may be near to $i$,
or nodes which are scattered through the network and far away from
$i$.
\end{abstract}

\pacs{05.10.-a, 05.40Fb, 89.70.Hj, 89.75.k}
\maketitle

\vspace{0.5cm}
\emph{`That which is static and repetitive is boring. That which
is dynamic and random is confusing.  In between lies art.'
(J. A. Locke)}

\section{Introduction} 

One special kind of dynamics regards the synchronization of events,
manifesting itself in many natural systems such as brain activity,
heart cell oscillations, and animal movement and development.
Synchronization is also closely related to the \emph{efficiency} of
several types of dynamics.  For instance, the uniform rate at which
components arrive at a factory is essential for avoiding delays or
accumulation of parts, the rate in which molecules are produced and
transformed into other molecules and tissues are critical for life,
among other possibilities.  Therefore, it is hardly surprising that
growing attention has been focused on the study of synchronization in
complex networks
(e.g.~\cite{Boccaletti:2006,Hong_etal:2004,Lee:2005,Boccaletti:2005,
Zhou:2006, Arenas:2006, Boccaletti:2007, Lodato:2007, Takashi:2006,
Sorrentino:2007, Ott:2007, Almendral:2007}).  However, the majority of
such investigations have addressed synchronization of linear dynamics.

The current work addresses the stochastic synchronization in transient
non-linear dynamics defined by self-avoiding random walks (the
extension to other types of walks is immediate).  We start by
estimating the transition probabilities of agents moving into a node
after having initiated random walks at the other nodes $h$ steps
before.  Several types of mean frequency and period of accesses are
defined, and the synchronizability is defined in terms of the
normalized entropy of the mean period of accesses to each node $i$ by
moving agents which initiated their walks at any of the $H$ previous
steps.  The concept of synchronous support of each node $i$ is
introduced, corresponding to the set of nodes which contribute the
most to the synchronizability of $i$.  The potential of these concepts
and methods for characterization of the transient, non-linear dynamics
in complex networks is illustrated with respect to 6 networks of
different types, leading to a series of interesting results regarding
not only the intrinsic overall synchronazibility of each type of
network, but also concerning correlations between synchronizability
and degree or outward accessibility.  Of particular interest has been
the identification of substantially lower synchronizabilities for the
Watts-Strogatz and path-regular models and the possibility of
relatively accurate estimation of the synchronizability of nodes in ER
networks from the respective degree.

This article starts by summarizing the main concepts in complex
networks and random walks and proceeds by defining synchronizability
and synchronous support.  These concepts are then applied for the
characterization of transient non-linear dynamics in 6 networks of
distinct kinds.

\section{Basic Concepts}

This section summarizes the main concepts in network representation,
measurement, random walks, as well as the six network models assumed
in the present article.

\subsection{Complex Networks Basics and Models}

A directed and weighted network $\Gamma$ can be represented in terms
of its \emph{weight matrix} $W$, so that each connection with weight
$r$ from node $i$ to $j$, $i,j \in \{1, 2, \ldots,N\}$ and $i \ne j$,
implies $W(j,i)=r$.  Observe that these weighs can correspond to
probabilities in some cases.  The absence of a connection between
those nodes is represented as $W(j,i)=0$.  $W$ becomes symmetric in
the case of undirected networks.  The topology of $\Gamma$ can be
represented in terms of the respective \emph{adjacency matrix} $K$,
obtained as $K(i,j) = \delta(W(i,j),1)$, where $\delta(a,b)$ is the
Kronecker's delta. Observe that $K$ is a binary matrix.  The
\emph{instrength} of a node $i$ is given by adding the weights of all
the respective incoming edges (an analogue definition holds for the
\emph{outstrength}). The indegree of a node $i$ in $K$ is the number
of incoming edges (similar for the outdegree).  If $K$ is symmetric,
$indegree = outdegree = degree$.

The \emph{immediate neighbors} of a node $i$ are those nodes which can
be reached from $i$ through a single edge.  A \emph{walk} can be
defined as a sequence of adjacent edges. A \emph{path} is a walk which
never repeats an edge or node.  The \emph{length} of a walk (or path)
is equal to the number of edges along it.  A \emph{random walk} is
defined by a moving agent as it performs a walk along the network.  A
emph{self-avoiding random walk} or \emph{path-walk} is defined by a
moving agent as it performs a walk without repeating nodes or edges.

Six models of complex networks are considered in the present article:
Erd\H{o}s-R\'enyi (ER), Barab\'asi-Albert (BA), Watts-Strogatz (WS), a
geographical model (GG), as well as two knitted networks (PN and
PA)~\cite{Albert_Barab:2002, Newman:2003, Dorogov_Mendes:2002,
Costa_surv:2007, Costa_path:2007, Costa_comp:2007,
Costa_longest:2007}.  The ER network (see also~\cite{Flory}) is
obtained by establishing connections between pairs of nodes with fixed
probability.  The traditional preferential attachment rule
(e.g.~\cite{Albert_Barab:2002, Newman:2003, Dorogov_Mendes:2002,
Costa_surv:2007}) is used here in order to obtain BA networks.  The
geographical network adopted in this article is obtained by
distributing $N$ nodes along a two-dimensional space and connecting
every pair of node whose nodes are closer than a given threshold.
\emph{Path-regular network} (PN) can be obtained by performing several 
path-walks along the $N$ nodes~\cite{Costa_path:2007, Costa_comp:2007,
Costa_longest:2007}, and \emph{path-transformed BA networks} are
obtained by converting a BA networks through the star-path
transformation of network connectivity~\cite{Costa_path:2007,
Costa_comp:2007}.  Therefore, this type of network includes paths with
several lengths (power law).  All networks considered in this work,
except the virtual networks established by the transition
probabilities, are undirected and have similar node degree $\left< k
\right> \approx 6$ and number of nodes $N$.  Only the largest
connected component, which contains most nodes because of the
relatively high adopted value of $\left< k \right>$, of each networks
is used herein.

\subsection{Synchronization in Random Walk Models}

\begin{figure}[htb]
  \vspace{0.3cm} \begin{center}
  \includegraphics[width=1\linewidth]{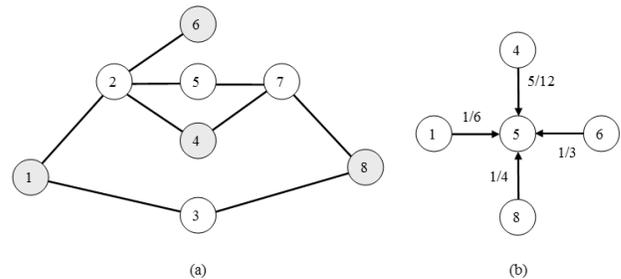} 
  \caption{A simple network containing $N=7$ nodes and the
           respective transition probabilities $P_2(q,5)$,
           with $q \in \Phi_2(5) = \{1, 4, 6, 8 \}$.
  }~\label{fig:ex_ps} \end{center}
\end{figure}

Let $\Gamma$ be a complex network in which moving agents perform
self-avoiding walks after having started at a specific node.  Let the
transition probabilities of the moving agent from node $i$ to node $j$
after $h$ steps along its self-avoiding walk to be $P_h(i,j)$.
Observe that, except for $h=1$, non-zero values of $P_h(i,j)$ do not
necessarily means the presence of a physical edge extending from node
$i$ to $j$, being here understood as \emph{virtual
edges}~cite{Costa:2004}.  The set of nodes $j$ which connect to node
$i$ through paths of length $h$ are henceforth represented as
$\Phi_h(i)$.  Observe that such a formulation can be immediately
extended to other types of Markovian walks.  The transition
probabilities $P_h(i,j)$ can be estimated computationally by
performing several walks of length $H$ starting at all the network
nodes and accumulating the number of visits to each node (the
probabilities are estimated by the relative frequencies).
Figure~\ref{fig:ex_ps}(b) illustrates the probabilities $P_2(q,5)$
considering the simple network in (a), $q \in \Phi_2(5) =
\{1,4,6,8\}$.  The nodes in gray correspond to the  nodes in 
$\Phi_2(5)$.  Having started the walk at node 1, the moving agent has
probability $P_1(a,2)=1/2$ of going to node 2 and then $P_1(2,5)=1/3$
of reaching node 5.  As this is the only path from 1 to 5, we have
$P_2(1,5)=P_1(1,2)P_1(2,5) = 1/6$ because of the Markovianity of such
a dynamic stochastic system.

\begin{table*}
\begin{tabular}{||c|c||}  \hline  
    Mean frequency of accesses to node $j$ &  \\ 
        by moving agents which,  $h$ steps ago,   &
        $f_h(i,j) = P_h(i,j)$     \\ 
        left from node $i$:  &  \\ \hline
    Mean frequency of accesses to node $i$ & \\ 
       by moving agents which, $h$ steps ago, & 
       $f_h(i) = \sum_{v=1}^{N} P_h(v,i)$   \\  
       left from all other nodes: & \\ \hline
    Mean frequency of accesses to node $j$ & \\ 
       by moving agents which, from 1 to $H$ steps ago,  &
       $f(i,j) = \sum_{h=1}^{H} P_h(i,j)$   \\  
       left from node $i$: &  \\   \hline
    Mean frequency of accesses to node $i$ & \\ 
       by moving agents which, from 1 to $H$ steps ago, &
       $f(i) = \sum_{h=1}^{H} \sum_{v=1}^{N} P_h(i,j) = 
             \sum_{v=1}^{N} f(v,i)$  \\  
       left from all other nodes: &  \\ \hline
\end{tabular}
\caption{The definition of the several mean frequency of accesses
         by moving agents.         
        }\label{tab:eqs}
\end{table*}

Observe that the transition probability matrix $P_h$ can be understood
as the transpose of a weight matrix describing virtual edges between
the original nodes.  After estimating $P_h(i,j)$ for $i,j \in \{1, 2,
\ldots, N\}$, it becomes possible to define the accessibility to each 
node~\cite{Costa_access:2008}.  The \emph{inward accessibility} of a
node $i$ after $h$ steps has been understood~\cite{Costa_access:2008}
as being equal to $IA_h(i) = \sum_{v=1}^{N} P_h(v,i)/(N-1)$.  The
outward accessibility of node $i$ after $h$ steps has been defined as
$OA_h(i) = exp(E_h(i))/(N-1)$, where $E(h(i)$ is the diversity
entropy~\cite{Costa_access:2008} of node $i$, given as $E_h(i) =
-\sum_{v=1}^{N} P_h(i,v)log(P_h(i,v))$.

It is assumed henceforth that self-avoiding walks are initiated from
all nodes at each time step.  The \emph{mean frequency of accesses} to
node $i$ received from node $j$ after $h$ steps is immediately given
as $f_h(j,i) = P_h(j,i)$.  The respective \emph{mean period of
accesses} is $T_h(j,i)=1/f_h(j,i)$. Observe that $T_h(i,j) \in \{1, 2,
\ldots, N-1 \}$ for any $i$ and $j$.  Table~\ref{tab:eqs} includes 
other important mean frequencies of accesses.  Observe that the
respective \emph{mean periods of accesses} can be immediately obtained
by taking the reciprocal of the respective mean frequencies.

It can be easily shown that $1 < T(i) \leq N-1$.  The probability that
all the moving agents which started their walks at the nodes in
$\Phi_h(i)$ reach node $i$ after $h$ steps is given as $S_h(i) =
\prod_{v \in \Phi_h(i)} P_h(v,i)$.  Consequently, the maximal surge of 
accesses to a node $i$ by nodes which started their walks at $h$ steps
before will occur with period $R_h(i) = 1/S_h(i)$.  The surge
frequency considering all time steps is $S_h(i) = \sum_{h=1}^{H}
\prod_{v \in \Phi_h(i)} P_h(v,i)$.

\begin{figure}[htb]
  \vspace{0.3cm} \begin{center}
  \includegraphics[width=1\linewidth]{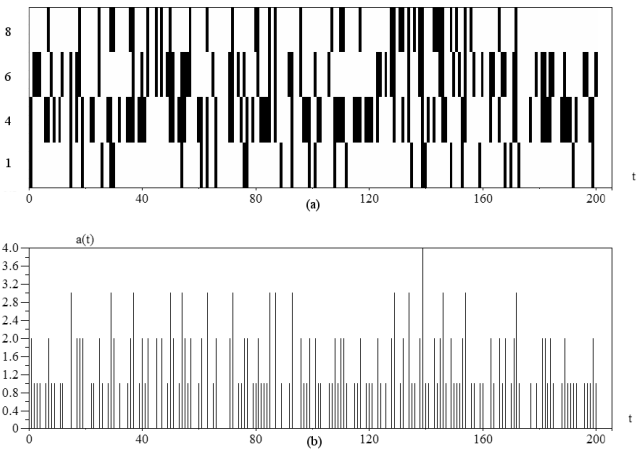} 
  \caption{The accesses to node 5 along time $t$ by agents which 
           left nodes 1, 4, 6 and 8 $h=2$ steps before (a), and 
           the respective total of such moving agents arriving 
           at node 5 at each time $t$.
           }~\label{fig:seq_a} \end{center}
\end{figure}

Figure~\ref{fig:seq_a} illustrates the above concepts with respect to
the network in Figure~\ref{fig:ex_ps}.  In Fig.~\ref{fig:seq_a}(a) are
shown the visits to node $5$ along the time $t$ received by moving
agents which started their walks at nodes 1, 4, 6 and 8 (each line of
the diagram).  Figure~\ref{fig:seq_a}(b) illustrates the total of
number of received visits at each time $t$ by agents which started
their walks $h=2$ steps before.  For this case, we have that
$f_2(1,5)=1/6$, $f_2(4,5)=5/12$, $f_2(6,5)=1/3$, $f_2(8,5)=1/4$, and
$f_2(i)=7/6$.

Though the agents arriving at node 5 after $2$ steps from the start of
their respective walks at the nodes $v \in \Phi_2(5)$ will present a
well-defined mean frequency of arrivals $f_2(v,5)$, because of the
statistical fluctuations it makes little sense speaking of
\emph{phase} or \emph{phase synchronization} between such time series.
Still, we consider in this work the synchronization between the agents
arriving at a node $i$ in terms of the similarity between their mean
frequencies of arrival at $i$.  In order to do so, let us assume that
each node $i$ needs to receive one visit from each of all other nodes
(the involving subset of these nodes are immediate) before being able
to treat them (e.g. to produce a respective molecule of perform some
computation).  Let us also suppose that, having received such visits,
node $i$ takes a period of time $\tau(i)$ to process them.  The ideal
situation regarding the processing speed and occupation of node $i$
takes place if and only $\tau(i) = T(i)$.  Such a matching of
frequencies is what we understand by \emph{synchronization} in the
present work.  It becomes interesting to devise an objective
quantification of the frequency synchronizability of each node in the
network.

\begin{figure}[htb]
  \vspace{0.3cm} 
  \begin{center}
  \includegraphics[width=0.5\linewidth]{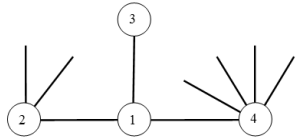} 
  \caption{A simple situation illustrating how the entropy of mean
     periods of accesses to a node (1) increases when the immediate
     neighbors of that node have different degrees.}~\label{fig:ex_sync} 
  \end{center}
\end{figure}

In this work, the \emph{synchronizability} is simply expressed in
terms of the normalized entropy of the effective periods of accesses
to each node per unit of time $T(i)$, i.e.

\begin{eqnarray}
  \epsilon(i) = - \sum_{v=1}^{N} p(T(v,i) log(p(T(v,i)))  \label{eq:entr}  \\
  \sigma(i) =  \frac{log(N-1) - \epsilon(i)}{log(N-1)} \label{eq:sync}  
\end{eqnarray}

where $T(v,i) = 1/f(v,i)$ (see Table~\ref{tab:eqs}), $\epsilon(i)$ is
the entropy of $T(v,i)$ and $log(N-1)$ is the maximum entropy possibly
achieved in any network.  It can be verified that $0 < \sigma(i)
\leq 1$. Thus, in case a node $i$ receives accesses with identical
frequencies $F(i)$, we will have $\sigma(i)=1$, indicating maximum
synchronizability.  The synchronizability of node $i$, and therefore
the efficiency in the use of the access received at that node, will
decrease when the probabilities $p(T(i))$ tend to concentrate on a few
nodes.  Figure~\ref{fig:ex_sync} illustrates that when a node $i$ (in
this case, node 2) has imediate neighbors with different degrees,
different rates of accesses to node $i$ (i.e. $P_1(2,1)=1/3$,
$P_1(3,1)=1$ and$P_1(4,1)=1/5$ ) will be obtained, implying high
entropy of mean periods of accesses to $i$.  In networks where a node
with high degree has an inherent change of being attached to nodes
with diverse degrees, a negative correlation between the
synchronizability and the node degree is expected.  Such a
relationship, however, is not guaranteed because the synchronizability
involves more global connectivity around each node.

It is also interesting to identify for each node $i$ what are the
other nodes in the network which contribute the most for the
synchronization of accesses to $i$.  This can be done by identifying
the highest value of $p(T(v,i))$, henceforth represented as
$p_{max}(i)$, and finding out which nodes $s$ have $T(s,i)$ nearly
equal to the respective mean period.  Such more highly synchronized
nodes are henceforth called the \emph{synchronous support} of node
$i$, henceforth represented as $S(i)$.  Finally, observe that the
probabilities $p(T(v,i))$ can be easily estimated from the histogram
of relative frequencies of $T(i)$, derived from the transition
matrices $P_h$.

\section{Results and Discussion}

In order to illustrate the concepts described in this article, we
considered 6 theoretical complex networks models --- namely
Erd\H{o}s-R\'enyi (ER), Barab\'asi-Albert (BA), Watts-Strogatz (WS), a
geographical model (GG) as well as two knitted networks (PN and PA),
all with $N=100$ (except GG, for which $N=91$) and $\left< k \right>
\approx 6$.  In all cases, a total of 2000 self-avoiding walks were
initiated at each of the nodes in order to estimate the transition
probabilities $P_h(i,j)$ for all nodes and for $h = 1$ to $H=10$.

\begin{figure*}[htb]
  \vspace{0.3cm} 
  \begin{center}
  \includegraphics[width=1\linewidth]{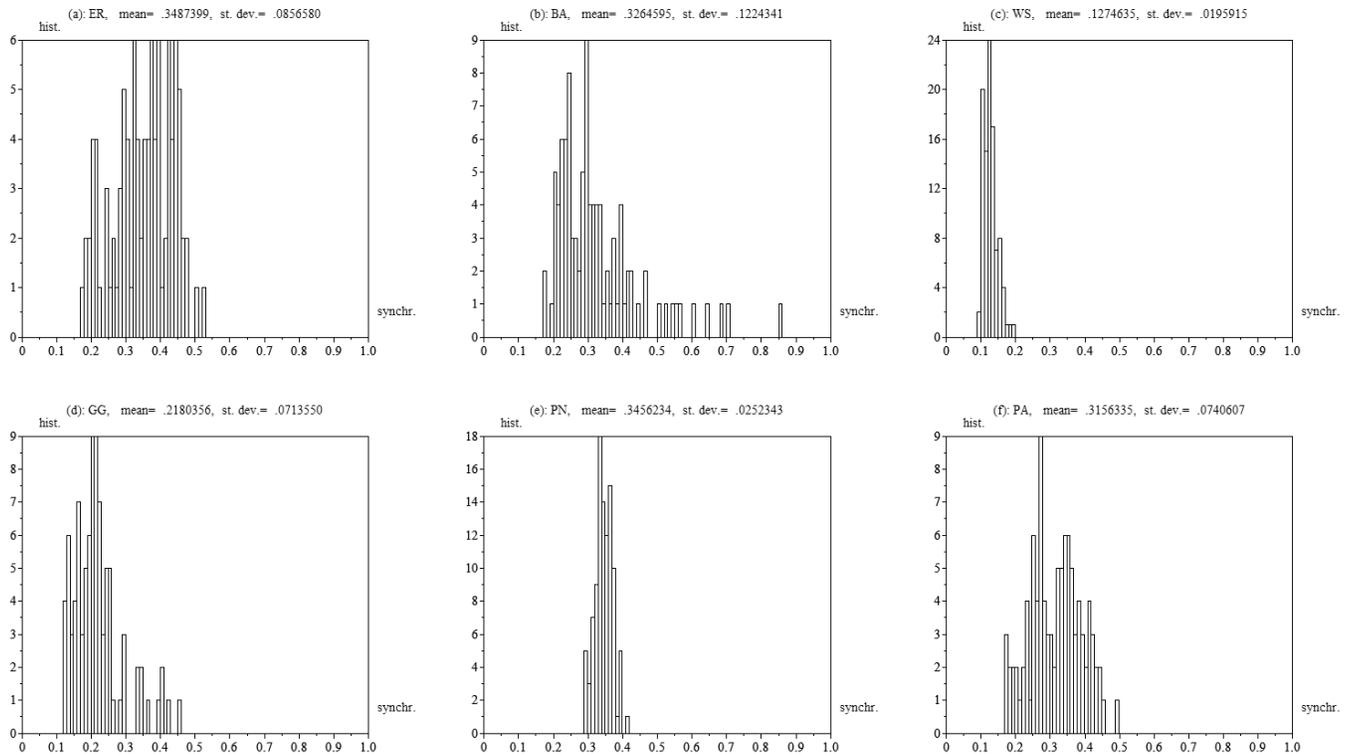} 
  \caption{Histograms of the synchronizability obtained for each
           of the 6 networks.}~\label{fig:sync_theo} 
  \end{center}
\end{figure*}

Figure~\ref{fig:sync_theo} shows the histograms of synchronizabilities
obtained considering each nodes for each network and $H=10$ (the mean
and standard deviations of the synchronizabilities are shown in the
respective titles). The ER, BA, PN and PA networks yielded similar and
relatively high synchronizatbilities, while the WS and GG were
characterized by substantially lower values (see
Figs.~\ref{fig:sync_theo}(c) and (d)).  The smallest dispersions of
synchronizability were obtained for WS and PN.  Additional experiments
performed indicated that the mean synchronizability tends to decrease
substantially with smaller values of $H$.

\begin{figure*}[htb]
  \vspace{0.3cm} 
  \begin{center}
  \includegraphics[width=1\linewidth]{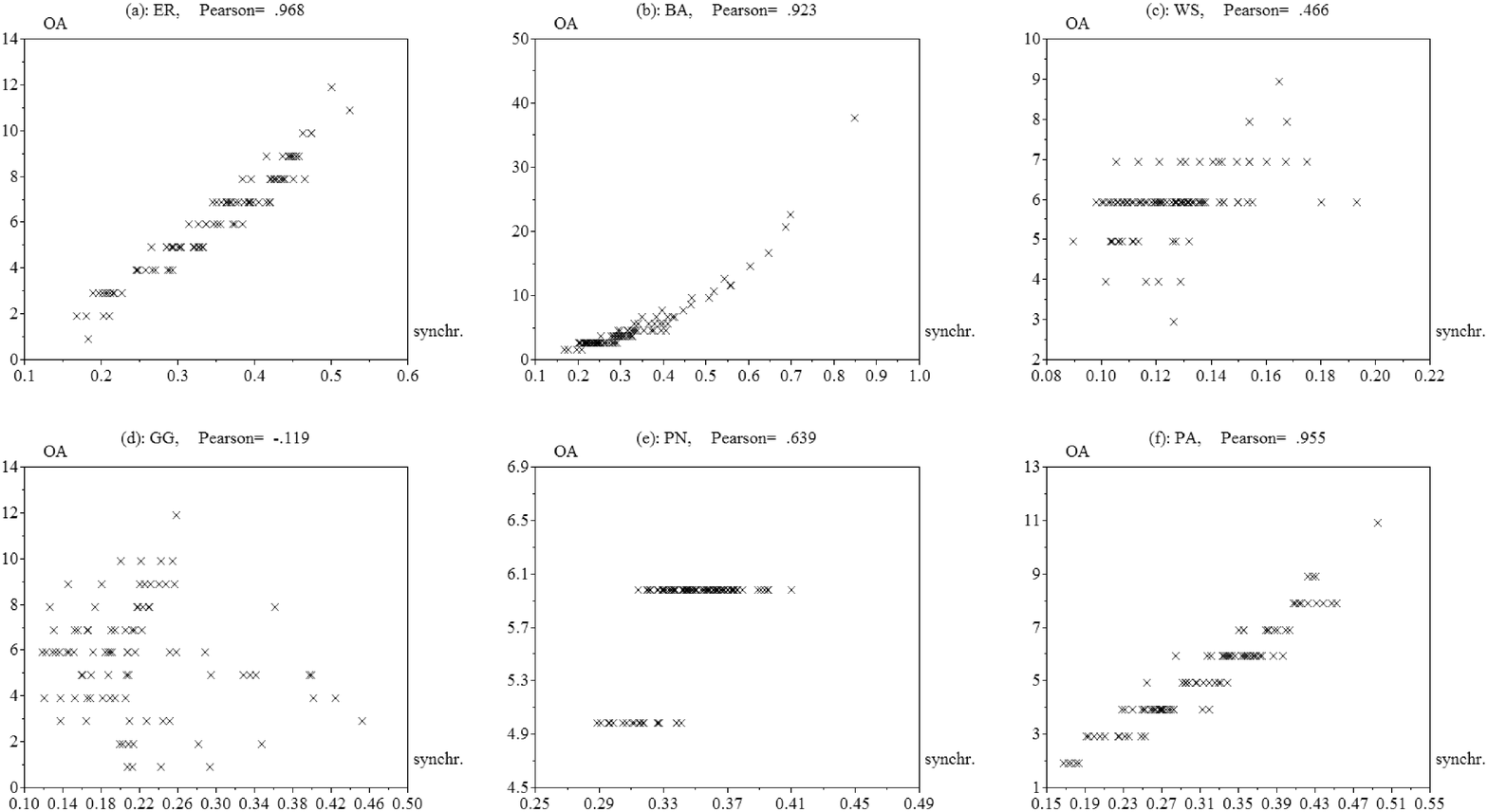} 
  \caption{Scatterplots of synchronizability against the node degree for
             the six considered networks.
          }~\label{fig:corrs_deg} 
  \end{center}
\end{figure*}

Figure~\ref{fig:corrs_deg} shows the scatterplots of the
synchronizabilities $\sigma$ of each node against its respective
degree $k$.  The Pearson correlation coefficients are shown in the
title of each plot. Such results show that the correlations between
$\sigma$ and $k$ vary markedly for each type of network.  ER showed
the highest correlation, suggesting that the synchronizability of its
nodes can be reasonably predicted from the respective degree.  In the
case of the BA structure, $\sigma$ exhibited an almost quadratic
increase with $k$, also leading to a well-defined relationship between
these two features.  Moderate positive correlations were also found
for the PN and PA network.  However, small correlations were obtained
for the WS and GG structures, with negative weak correlation being
observed for the WS network.  This implies that it is virtually to
infer the synchronizability of a node in the GG structure from its
degree.

\begin{figure*}[htb]
  \vspace{0.3cm} \begin{center}
  \includegraphics[width=0.7\linewidth]{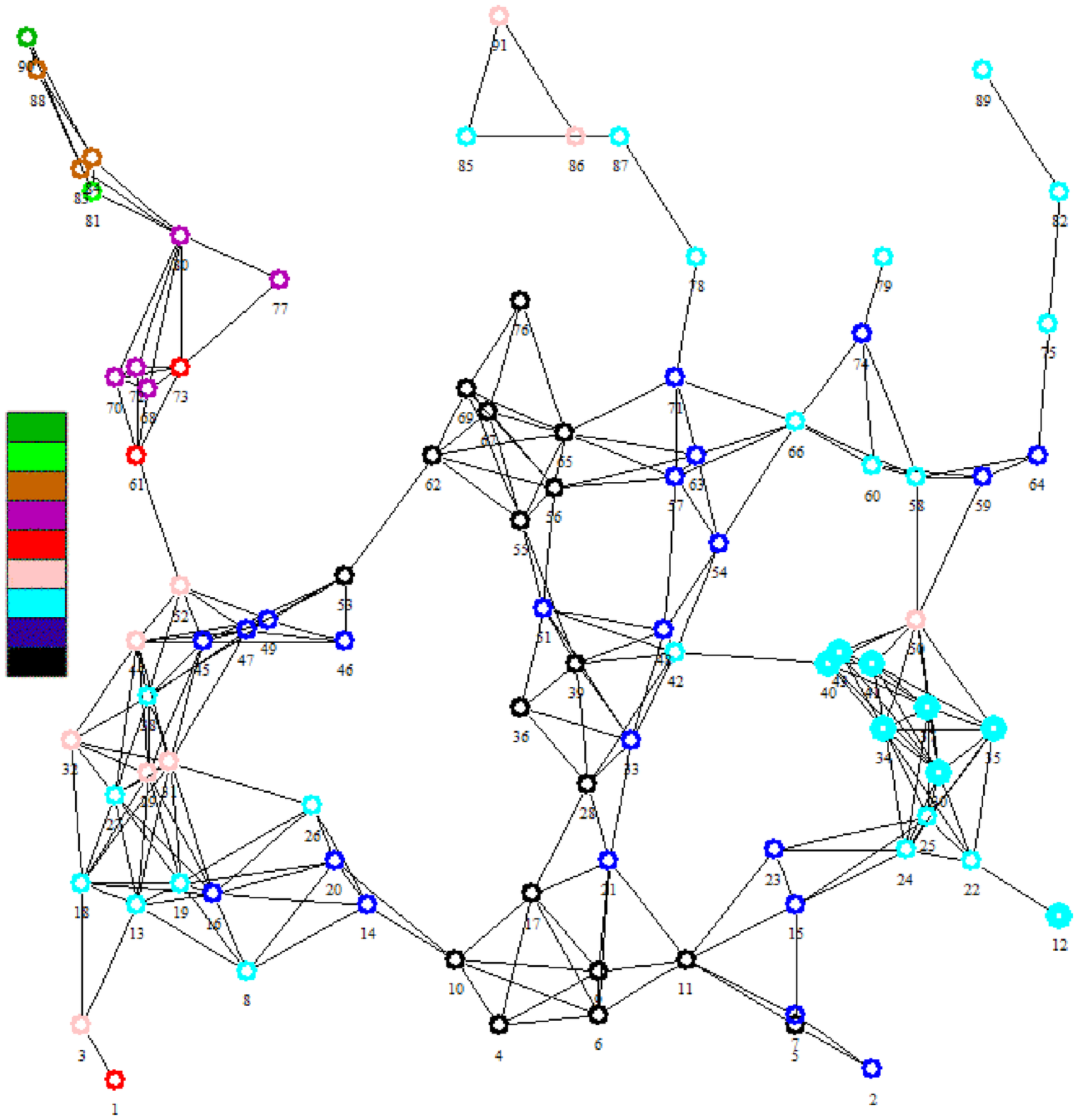} 
  \caption{The values of the synchronizability ($H=10$) of each 
           node in the GG network considered in this article are shown
           accordingly with the color scale (increasing from bottom
           upwards). The nodes belonging to the synchronous support 
           of node 50 are shown with wider borders.
           }~\label{fig:graph_GG} \end{center}
\end{figure*}

Figure~\ref{fig:graph_GG} shows the geographical network considered in
this article, with the synchronizabilities of each node identified by
the color legend.  Node 90 presented the highest synchronizability
value ($\sigma(90)=0.45$).  Observe that the values of $\sigma$ tend
to be similar among neighboring nodes (assortativeness of
synchronizability).  The more central nodes, identified by larger
outward accessibility, resulted with the lowest synchronizabilities.
Also, the more isolated group of nodes at the upper left-hand side of
the figure presented the highest values of synchronizability.  Because
such nodes are known~\cite{Costa_access:2008} to have low outward
accessibility $OA$, it is interesting to consider possible
correlations between this measurement and synchronizability.

\begin{figure*}[htb]
  \vspace{0.3cm} 
  \begin{center}
  \includegraphics[width=1\linewidth]{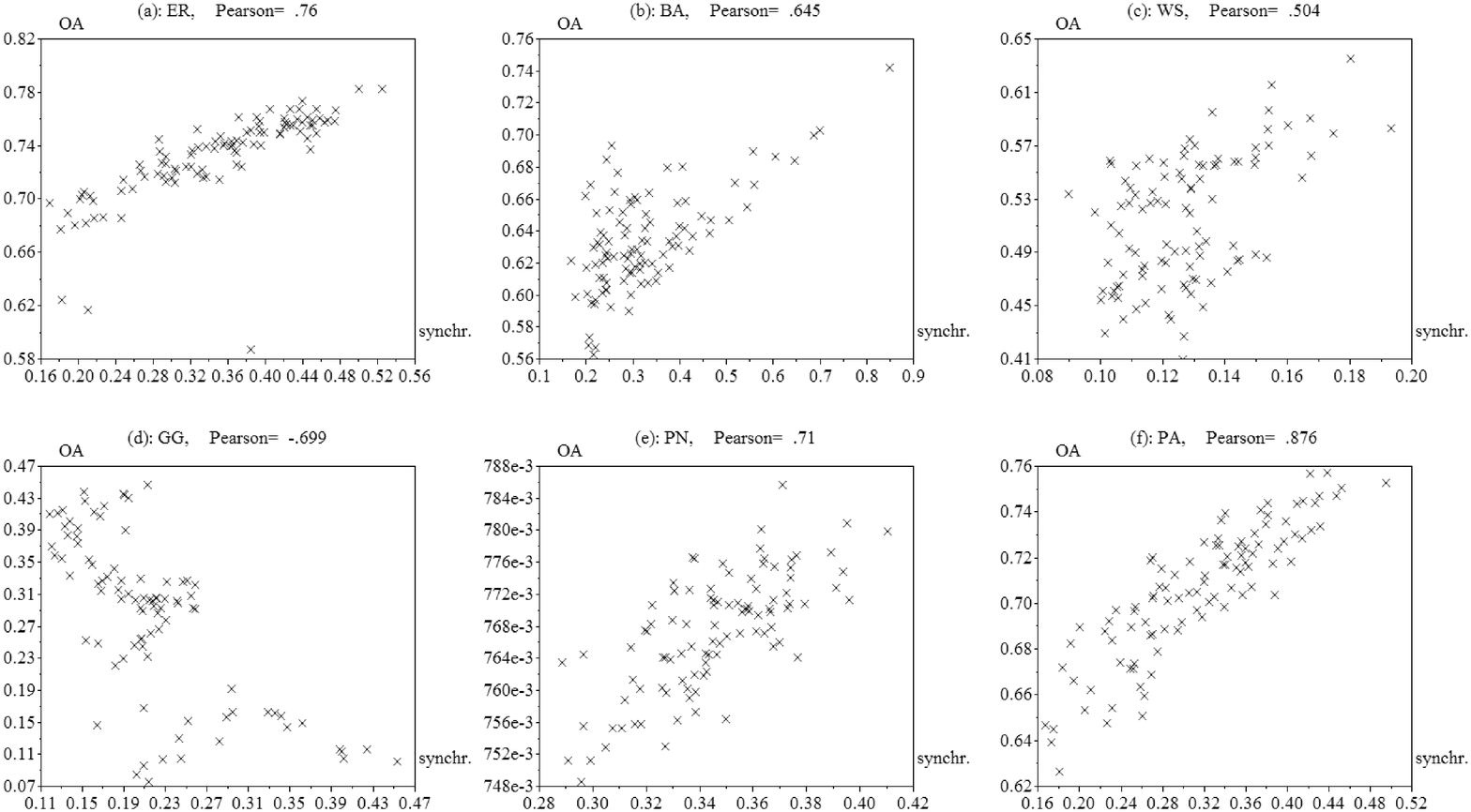} 
  \caption{Scatterplots of the synchronizability $\sigma$ against
           the respective node degrees $k$ obtained for each of the
           6 networks.  
           }~\label{fig:corrs_OA} 
  \end{center}
\end{figure*}

Figure~\ref{fig:corrs_OA} shows the scatterplots obtained considering
the synchronizabilities and outward accessibilities (corresponding to
the mean value of the accessibilities along the $H$ steps) of each
node for all the networks in this article.  The Pearson correlation
coefficients are also shown respectively to each plot.  The results
show that varying correlations are verified between $\sigma$ and $OA$
for each type of networl.  The ER and PA structures showed
particularly strong positive correlations (Fig.~\ref{fig:corrs_OA}a
and f), while the WS model exhibited the weakest correlation
(Fig.~\ref{fig:corrs_OA}c).  Interestingly, the synchronizability
resulted negatively correlated for the GG network
(Fig.~\ref{fig:corrs_OA}d).

\begin{figure}[htb]
  \vspace{0.3cm} 
  \begin{center}
  \includegraphics[width=1\linewidth]{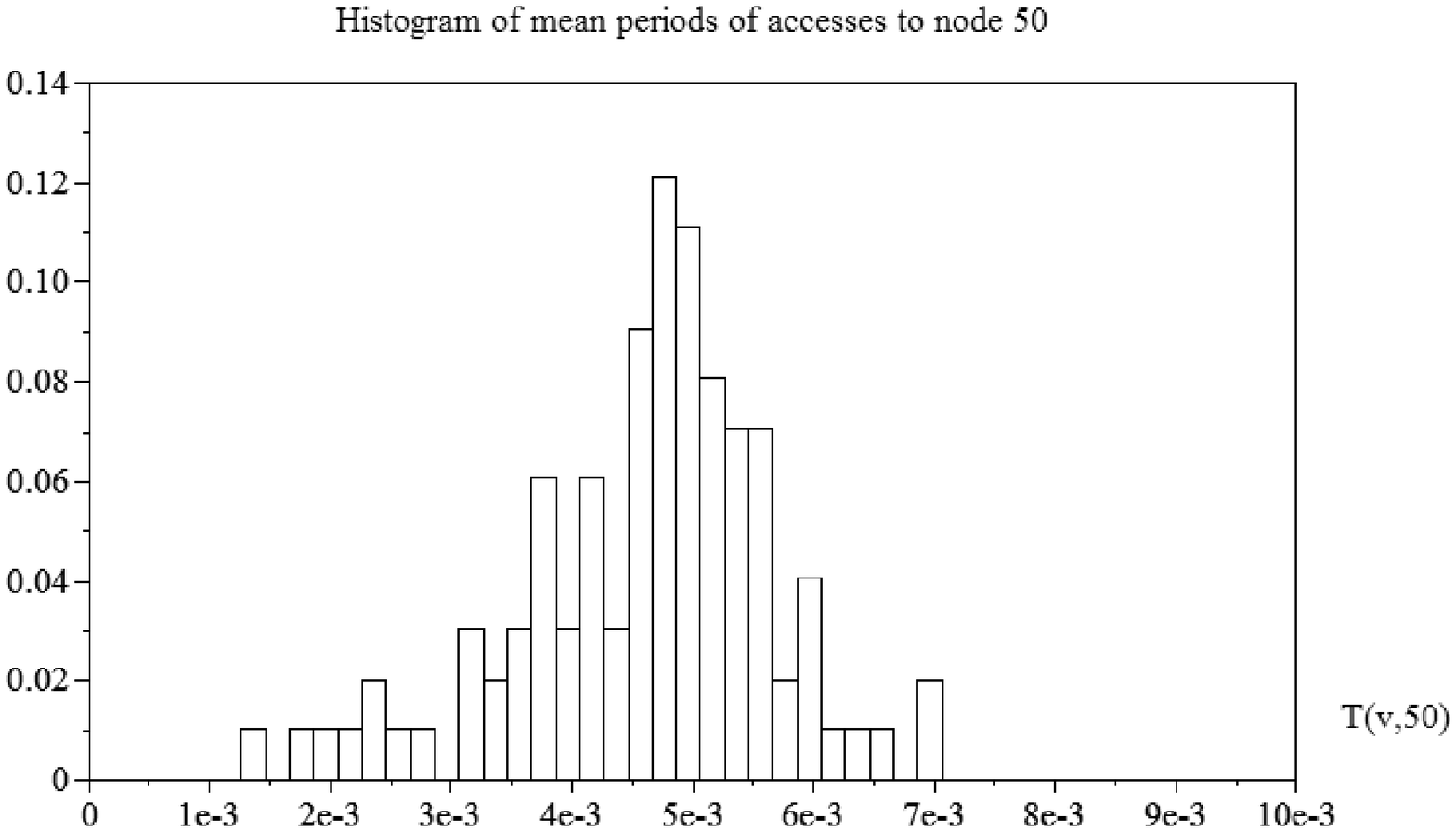} 
  \caption{Histogram of the relative frequencies of mean periods 
              of accesses to node 50 from all other nodes in
              the network.}~\label{fig:hist_T_ref} 
  \end{center}
\end{figure}

The nodes belonging to the synchronous support of node 50, which has
an intermediate synchronizability of $\sigma(50) = 0.26$, are shown in
Figure~\ref{fig:graph_GG} with wider borders.  The respective
histogram of relative frequencies of mean periods of accesses to this
node, i.e. $T(v,50)$, $v=1, 2, \ldots, N$, is depicted in
Figure~\ref{fig:hist_T_ref}.  Observe that the synchronous support of
node 50 comprises most of the nodes in the small community to which it
belongs, except only for nodes 22, 24 and 25.  Similar relationship
with communities has been verified for nodes 61 and 52.  However, the
synchronous support of node 1 was found to include nodes 2, 15, 25,
36, 48, 63 and 71, which are all far away and scattered through the
right-had side of the network.  This suggests that the synchronous
supports of nodes in a GG network do not seem to follow a typical
structural pattern.

\section{Concluding Remarks}

We have considered the situation where, after leaving continuously
from each of the nodes in a complex networks, moving agents perform
self-avoiding walks of fixed and finite length along complex networks.
After estimating the probabilities of accesses received by each node
by agents leaving from each of the other nodes a number of steps
behind, it is possible to consider synchronization between the
respective mean frequencies.  An objective measurement of the
synchronizability of each node has been reported which takes into
account the entropies of the mean periods of accesses from all other
nodes by moving agents which initiated their walks at any of the $H$
previous steps.  It has been shown that such a kind of synchronization
optimizes the deliver to the reference node of agents originating at
each of the other nodes at a constant rate and well-defined, avoiding
accumulation or lack of accesses along the critical period.  Such a
result also allows the identification of optimal processing times for
each node.  Modifications of the reported framework to consider
different rates of random walks emanating from each node or accesses
from subsets of the network nodes, instead of from all nodes, are
straightforward and can be useful for more general applications.  The
concept of synchronous support of a node has also been suggested in
order to identify the nodes which most contribute to the
synchronizability of each node.  Several interesting results have been
obtained, including the identification of distinct overall
synchronizability exhibited among 6 types of networks (with the WS and
GG being less synchronizable), as well as the identification, in the
case of the GG, of assortativeness of synchronizability and the fact
that the synchronous support of the nodes tend not to exhibit a
typical pattern.  The correlations between the synchronizability and
degree or outward accessibility have also been considered and found to
present diverse patterns.  For instance, while the synchronizability
of the nodes in an ER structure can be reasonably inferred from their
respective nodes with relatively high accuracy, it is virtually
impossible to perform such predictions in the case of the GG
structure.  Therefore, interesting patterns of relationships have been
identified between a topological feature (node degree) and the dynamic
property of synchronizability in the transient non-linear dynamics
defined by self-avoiding random walks.

Future developments include the search for additional possible
topological features which can explain or predict synchronizability,
studies of scaling with the average degree and network size, and the
investigation of synchronizabilities implied by different types of
walks.  Also interesting would be to search further for possible
relationships between the synchronizability and community structure.
In addition, the concepts described in this article have good
potential for applications in real-world problems, including
biological systems, transportation and communications networks.  Of
special interest is the study of non-linear transient sychronizability
in neuronal or cortical networks, especially by considering frontwave
random walks, i.e. the progression of accesses along the successive
neighborhoods of each node (e.g.~\cite{Costa:2004}).

\begin{acknowledgments}
Luciano da F. Costa thanks CNPq (308231/03-1) and FAPESP (05/00587-5)
for sponsorship.
\end{acknowledgments}

\bibliography{nlin_sync}

\begin{thebibliography}{22}
\expandafter\ifx\csname natexlab\endcsname\relax\def\natexlab#1{#1}\fi
\expandafter\ifx\csname bibnamefont\endcsname\relax
  \def\bibnamefont#1{#1}\fi
\expandafter\ifx\csname bibfnamefont\endcsname\relax
  \def\bibfnamefont#1{#1}\fi
\expandafter\ifx\csname citenamefont\endcsname\relax
  \def\citenamefont#1{#1}\fi
\expandafter\ifx\csname url\endcsname\relax
  \def\url#1{\texttt{#1}}\fi
\expandafter\ifx\csname urlprefix\endcsname\relax\def\urlprefix{URL }\fi
\providecommand{\bibinfo}[2]{#2}
\providecommand{\eprint}[2][]{\url{#2}}

\bibitem[{\citenamefont{Boccaletti et~al.}(2006)\citenamefont{Boccaletti,
  Latora, Moreno, Chavez, and Hwang}}]{Boccaletti:2006}
\bibinfo{author}{\bibfnamefont{S.}~\bibnamefont{Boccaletti}},
  \bibinfo{author}{\bibfnamefont{V.}~\bibnamefont{Latora}},
  \bibinfo{author}{\bibfnamefont{Y.}~\bibnamefont{Moreno}},
  \bibinfo{author}{\bibfnamefont{M.}~\bibnamefont{Chavez}}, \bibnamefont{and}
  \bibinfo{author}{\bibfnamefont{D.}~\bibnamefont{Hwang}},
  \bibinfo{journal}{Phys. Rep.} \textbf{\bibinfo{volume}{424}},
  \bibinfo{pages}{175} (\bibinfo{year}{2006}).

\bibitem[{\citenamefont{Lee}(2005)}]{Lee:2005}
\bibinfo{author}{\bibfnamefont{D.~S.} \bibnamefont{Lee}},
  \bibinfo{journal}{Phys. Rev. E} \textbf{\bibinfo{volume}{72}},
  \bibinfo{pages}{026208} (\bibinfo{year}{2005}).

\bibitem[{\citenamefont{Zhou et~al.}(2006)\citenamefont{Zhou, Motter, and
  Kurths}}]{Zhou:2006}
\bibinfo{author}{\bibfnamefont{C.}~\bibnamefont{Zhou}},
  \bibinfo{author}{\bibfnamefont{A.~E.} \bibnamefont{Motter}},
  \bibnamefont{and} \bibinfo{author}{\bibfnamefont{J.}~\bibnamefont{Kurths}},
  \bibinfo{journal}{Phys. Rev. Letts.} \textbf{\bibinfo{volume}{96}},
  \bibinfo{pages}{034101} (\bibinfo{year}{2006}).

\bibitem[{\citenamefont{Arenas et~al.}(2006)\citenamefont{Arenas, Guilera, and
  Vicente}}]{Arenas:2006}
\bibinfo{author}{\bibfnamefont{A.}~\bibnamefont{Arenas}},
  \bibinfo{author}{\bibfnamefont{A.~D.} \bibnamefont{Guilera}},
  \bibnamefont{and} \bibinfo{author}{\bibfnamefont{C.~J.~P.}
  \bibnamefont{Vicente}}, \bibinfo{journal}{Phys. Rev. Letts.}
  \textbf{\bibinfo{volume}{96}}, \bibinfo{pages}{114102}
  (\bibinfo{year}{2006}).

\bibitem[{\citenamefont{Boccaletti et~al.}(2007)\citenamefont{Boccaletti,
  Ivachenko, Latora, Pluchino, and Rapisarda}}]{Boccaletti:2007}
\bibinfo{author}{\bibfnamefont{S.}~\bibnamefont{Boccaletti}},
  \bibinfo{author}{\bibfnamefont{M.}~\bibnamefont{Ivachenko}},
  \bibinfo{author}{\bibfnamefont{V.}~\bibnamefont{Latora}},
  \bibinfo{author}{\bibfnamefont{A.}~\bibnamefont{Pluchino}}, \bibnamefont{and}
  \bibinfo{author}{\bibfnamefont{A.}~\bibnamefont{Rapisarda}},
  \bibinfo{journal}{Phys. Rev. E} \textbf{\bibinfo{volume}{75}},
  \bibinfo{pages}{045102} (\bibinfo{year}{2007}).

\bibitem[{\citenamefont{Lodato et~al.}(2007)\citenamefont{Lodato, Boccaletti,
  and Latora}}]{Lodato:2007}
\bibinfo{author}{\bibfnamefont{I.}~\bibnamefont{Lodato}},
  \bibinfo{author}{\bibfnamefont{S.}~\bibnamefont{Boccaletti}},
  \bibnamefont{and} \bibinfo{author}{\bibfnamefont{V.}~\bibnamefont{Latora}},
  \bibinfo{journal}{Phys. Rev. Letts} \textbf{\bibinfo{volume}{78}},
  \bibinfo{pages}{28001} (\bibinfo{year}{2007}).

\bibitem[{\citenamefont{Nishiwaka and Motter}(2006)}]{Takashi:2006}
\bibinfo{author}{\bibfnamefont{T.}~\bibnamefont{Nishiwaka}} \bibnamefont{and}
  \bibinfo{author}{\bibfnamefont{A.~E.} \bibnamefont{Motter}},
  \bibinfo{journal}{Phys. D} \textbf{\bibinfo{volume}{224}},
  \bibinfo{pages}{77} (\bibinfo{year}{2006}).

\bibitem[{\citenamefont{Sorrentino et~al.}(2007)\citenamefont{Sorrentino,
  di~Bernardo, Garofalo, and Chen}}]{Sorrentino:2007}
\bibinfo{author}{\bibfnamefont{F.}~\bibnamefont{Sorrentino}},
  \bibinfo{author}{\bibfnamefont{M.}~\bibnamefont{di~Bernardo}},
  \bibinfo{author}{\bibfnamefont{F.}~\bibnamefont{Garofalo}}, \bibnamefont{and}
  \bibinfo{author}{\bibfnamefont{G.}~\bibnamefont{Chen}},
  \bibinfo{journal}{Phys. Rev. E} \textbf{\bibinfo{volume}{75}},
  \bibinfo{pages}{046103} (\bibinfo{year}{2007}).

\bibitem[{\citenamefont{Sorrentino and Ott}(2007)}]{Ott:2007}
\bibinfo{author}{\bibfnamefont{F.}~\bibnamefont{Sorrentino}} \bibnamefont{and}
  \bibinfo{author}{\bibfnamefont{E.}~\bibnamefont{Ott}},
  \bibinfo{journal}{Phys. Rev. E} \textbf{\bibinfo{volume}{76}},
  \bibinfo{pages}{056114} (\bibinfo{year}{2007}).

\bibitem[{\citenamefont{Almendral and Guilera}(2007)}]{Almendral:2007}
\bibinfo{author}{\bibfnamefont{J.~A.} \bibnamefont{Almendral}}
  \bibnamefont{and} \bibinfo{author}{\bibfnamefont{A.~D.}
  \bibnamefont{Guilera}} (\bibinfo{year}{2007}),
  \bibinfo{note}{arXiv:0705.3216}.

\bibitem[{\citenamefont{Hwang et~al.}(2005)\citenamefont{Hwang, Chavez, Amann,
  and Boccaletti}}]{Boccaletti:2005}
\bibinfo{author}{\bibfnamefont{D.~U.} \bibnamefont{Hwang}},
  \bibinfo{author}{\bibfnamefont{M.}~\bibnamefont{Chavez}},
  \bibinfo{author}{\bibfnamefont{A.}~\bibnamefont{Amann}}, \bibnamefont{and}
  \bibinfo{author}{\bibfnamefont{S.}~\bibnamefont{Boccaletti}},
  \bibinfo{journal}{Phys. Rev. letts.} \textbf{\bibinfo{volume}{94}},
  \bibinfo{pages}{138701} (\bibinfo{year}{2005}).

\bibitem[{\citenamefont{Hong et~al.}(2004)\citenamefont{Hong, Kim, Choi, and
  Park}}]{Hong_etal:2004}
\bibinfo{author}{\bibfnamefont{H.}~\bibnamefont{Hong}},
  \bibinfo{author}{\bibfnamefont{B.~J.} \bibnamefont{Kim}},
  \bibinfo{author}{\bibfnamefont{M.~Y.} \bibnamefont{Choi}}, \bibnamefont{and}
  \bibinfo{author}{\bibfnamefont{H.}~\bibnamefont{Park}},
  \bibinfo{journal}{Phys. Rev. E} \textbf{\bibinfo{volume}{69}},
  \bibinfo{pages}{067105} (\bibinfo{year}{2004}).

\bibitem[{\citenamefont{Albert and Barab\'asi}(2002)}]{Albert_Barab:2002}
\bibinfo{author}{\bibfnamefont{R.}~\bibnamefont{Albert}} \bibnamefont{and}
  \bibinfo{author}{\bibfnamefont{A.~L.} \bibnamefont{Barab\'asi}},
  \bibinfo{journal}{Rev. Mod. Phys.} \textbf{\bibinfo{volume}{74}},
  \bibinfo{pages}{47} (\bibinfo{year}{2002}).

\bibitem[{\citenamefont{Newman}(2003)}]{Newman:2003}
\bibinfo{author}{\bibfnamefont{M.~E.~J.} \bibnamefont{Newman}},
  \bibinfo{journal}{SIAM Rev.} \textbf{\bibinfo{volume}{45}},
  \bibinfo{pages}{167} (\bibinfo{year}{2003}).

\bibitem[{\citenamefont{Dorogovtsev and Mendes}(2002)}]{Dorogov_Mendes:2002}
\bibinfo{author}{\bibfnamefont{S.~N.} \bibnamefont{Dorogovtsev}}
  \bibnamefont{and} \bibinfo{author}{\bibfnamefont{J.~F.~F.}
  \bibnamefont{Mendes}}, \bibinfo{journal}{Advs. in Phys.}
  \textbf{\bibinfo{volume}{51}}, \bibinfo{pages}{1079} (\bibinfo{year}{2002}).

\bibitem[{\citenamefont{da~F.~Costa et~al.}(2007)\citenamefont{da~F.~Costa,
  Rodrigues, Travieso, and Boas}}]{Costa_surv:2007}
\bibinfo{author}{\bibfnamefont{L.}~\bibnamefont{da~F.~Costa}},
  \bibinfo{author}{\bibfnamefont{F.~A.} \bibnamefont{Rodrigues}},
  \bibinfo{author}{\bibfnamefont{G.}~\bibnamefont{Travieso}}, \bibnamefont{and}
  \bibinfo{author}{\bibfnamefont{P.~R.~V.} \bibnamefont{Boas}},
  \bibinfo{journal}{Advs. in Phys.} \textbf{\bibinfo{volume}{56}},
  \bibinfo{pages}{167} (\bibinfo{year}{2007}).

\bibitem[{\citenamefont{da~F.~Costa}(2007{\natexlab{a}})}]{Costa_path:2007}
\bibinfo{author}{\bibfnamefont{L.}~\bibnamefont{da~F.~Costa}}
  (\bibinfo{year}{2007}{\natexlab{a}}), \bibinfo{note}{arXiv:0711.1271}.

\bibitem[{\citenamefont{da~F.~Costa}(2007{\natexlab{b}})}]{Costa_comp:2007}
\bibinfo{author}{\bibfnamefont{L.}~\bibnamefont{da~F.~Costa}}
  (\bibinfo{year}{2007}{\natexlab{b}}), \bibinfo{note}{arXiv:0711.2736}.

\bibitem[{\citenamefont{da~F.~Costa}(2007{\natexlab{c}})}]{Costa_longest:2007}
\bibinfo{author}{\bibfnamefont{L.}~\bibnamefont{da~F.~Costa}}
  (\bibinfo{year}{2007}{\natexlab{c}}), \bibinfo{note}{arXiv:0712.0415}.

\bibitem[{\citenamefont{Flory}(1941)}]{Flory}
\bibinfo{author}{\bibfnamefont{P.~J.} \bibnamefont{Flory}},
  \bibinfo{journal}{Journal of the American Chemical Society}
  \textbf{\bibinfo{volume}{63}}, \bibinfo{pages}{3083} (\bibinfo{year}{1941}).

\bibitem[{\citenamefont{da~F.~Costa}(2007{\natexlab{d}})}]{Costa_access:2008}
\bibinfo{author}{\bibfnamefont{L.}~\bibnamefont{da~F.~Costa}}
  (\bibinfo{year}{2007}{\natexlab{d}}), \bibinfo{note}{arXiv:0801.1982}.

\bibitem[{\citenamefont{da~F.~Costa}(2004)}]{Costa:2004}
\bibinfo{author}{\bibfnamefont{L.}~\bibnamefont{da~F.~Costa}},
  \bibinfo{journal}{Phys. Rev. Lett.} \textbf{\bibinfo{volume}{93}},
  \bibinfo{pages}{098702} (\bibinfo{year}{2004}).

\end{thebibliography}
\end{document}